\documentclass[10pt,twoside]{hsqcd}
\usepackage{inputenc}
\usepackage{epsf,amsmath}
\usepackage{graphicx}

\begin{document}

\title{RENORMALIZATION IN EFFECTIVE THEORIES: \\
PRESCRIPTIONS  FOR KAON-NUCLEON RESONANCE PARAMETERS
\thanks{This work is supported by INTAS (project 587, 2000) and by
Ministry of Education of Russia (Programme ``Universities of Russia'').}
}

\author{ K.~M.~SEMENOV-TIAN-SHANSKY  \\
St. Petersburg State University \\
\\
E-mail: semenov@pdmi.ras.ru  }

\maketitle

\begin{abstract}
\noindent
In this talk we show how it is possible to apply the general scheme of
effective scattering theory to the description of hadronic processes.
We perform the numerical tests of the tree level bootstrap constraints
for renormalization prescriptions in the case of elastic kaon-nucleon
scattering process.
\end{abstract}



\markboth{\large \sl  K.M. Semenov-Tian-Shansky
\hspace*{2cm} HSQCD 2004}
{\large \sl \hspace*{1cm} RENORMALIZATION IN EFFECTIVE THEORIES...}

\section{Introduction}

In papers
\cite{AVVV}-\cite{Essential}
an attempt is made to develop an effective field theory formalism
suitable for description of hadronic scattering processes (see also
\cite{hsqcd}).
It was shown that the requirements of
consistency of perturbation series for scattering amplitude lead to
certain restrictions for the effective Hamiltonian parameters that
are called
{\em bootstrap equations}.
Actually, we are unable to solve the bootstrap system explicitly. So,
roughly speaking,  the only way to check the consistency in our
effective theory approach is the numerical testing.

An important property of the bootstrap system is its renormalization
invariance. This property allows one to compare with experiment the
results that follow already from the tree level bootstrap system.
In many cases this data fitting leads to reasonable consequences. This
can be considered as a strong evidence in favor of consistency of our
approach. The similar verification was successful in the cases of
$\pi K$
\cite{AVVV}
and
$\pi N$
elastic scattering processes (see references in
\cite{hsqcd}).

In this talk we discuss the application of our formalism to the case
 of
$KN$
elastic scattering. The resonance spectrum of
$KN$
reaction is measured with much less precision than that of
$\pi N$
reaction. However, it is possible to single out the set of sum
rules that are well saturated by the known experimental data.

On the other hand, those sum rules that are not so well saturated with
available data permit us to speculate about the possible scenarios
that allow to amend the situation. So here we also aim to show that
our approach is a powerful tool to study the resonance spectrum.

\section{Bootstrap for KN scattering}

The amplitude of
$KN$
elastic scattering
$M^{\ \  \beta j}_{\alpha i}=
\left\langle N_\beta(k') K_j(p')
\left|(S-1)\right| N_\alpha(k) K_i(p) \right\rangle$
can be presented in the following form:
$$
M_{\alpha i}^{\ \ \ \beta j}=\delta_\alpha^{\; \beta} \delta_i^{\; j}
M^{+} (\lambda, \lambda')+
\delta_\alpha^{ \; j} \delta_i^{\; \beta}
M^{-} (\lambda, \lambda'),
$$
where
$
M^{\pm }(\lambda, \lambda',s,t,u)=\overline{u}^{+}(\lambda',k')
\left\{
A^\pm(s,t,u)+\frac{\hat{p}+\hat{p}'}{2}B^\pm(s,t,u)
\right\}
u^{-}(\lambda,k).
$
Here
$k,k'$ $(p,p')$
stand for the nucleon (kaon) momenta,
$\hat{p} \equiv p_\mu \gamma^\mu$;
$ \alpha, i, \beta, j= 1,2$
are the isospin indices;
$\lambda, \lambda'$ stand
for nucleon spin variables;
$\overline{u}(k',\lambda')$, $u(k,\lambda)$ ---
for Dirac spinors. Invariant amplitudes
$A^{\pm}$
and
$B^{\pm}$
are the functions of an arbitrary pair of Mandelstam kinematical
variables
$s,t,u$.

The detailed theoretical background of our calculations is discussed in
\cite{hsqcd}.
Here we shall only briefly recall the main steps needed to construct
the set of tree level bootstrap constraints for renormalization
prescriptions (RP's) in
$KN$
reaction.

We work in the framework of the general formalism of effective
theories. This means that the corresponding interaction Hamiltonian
contains all local terms consistent with given algebraic symmetry
requirements. We consider a very narrow class of so-called
{\em localizable}
effective theories. In this case to construct a consistent tree level
approximation it is necessary to turn to the
{\em extended perturbation scheme}
which, along with the fields of stable particles, also contains an
infinite number of fields corresponding to auxiliary unstable ones
(resonances) of arbitrary high spin and mass. The tree level amplitude
of a scattering process
$2 \rightarrow 2$
calculated in this formalism takes a form of an infinite sum of
resonance exchange graphs plus another (also infinite) sum of all
possible contact terms. Thus one needs to establish certain guiding
principle that would allow to fix the order of summation of this
formal series for tree level amplitude. This problem can be solved by
passing to the
{\em minimal parametrization}
(see
\cite{Essential})
and by using the method of Cauchy forms. Minimal parametrization
allows one to get rid of those combination of Hamiltonian parameters
which do not contribute to the renormalized
$S$-matrix.
It can be shown that the tree level amplitude is completely determined
by the values of three-leg minimal vertices (in some cases, one also
needs to impose one additional RP fixing the value of the amplitude at
certain kinematical point).

The method of Cauchy forms allows one to present the tree-level
$2 \rightarrow 2$
scattering amplitude as a uniformly convergent series of pole
contributions in three mutually intersecting (near the corners of
Mandelstam triangle) layers
$
B_s \{ s \sim 0 \} , \, B_t \{ t \sim 0 \}, \, B_u \{ u \sim 0 \}.
$
Bootstrap system naturally arises as a requirement that the Cauchy
forms (different in different layers) should coincide in the domains
of intersection of layers. For example, let us consider the system of
those tree level bootstrap constraints for
$A^-$
invariant amplitude which appear from the domain where the layers
$B_s$
and
$B_u$
intersect. Namely, the difference of Cauchy forms in two layers
$
\left. \widetilde{A}^-(s,u) \right|_{B_s}-\left.
  \widetilde{A}^-(u,s) \right|_{B_u} \equiv \Phi^-_A(u,s)
$
should be identically zero in the vicinity of the point
$s=0, \,u=0$:
\begin{equation}
\left.
  \frac{\partial^{p+k}}{\partial u^k  \partial s^p} \Phi_A^-(u,s)
 \right|_{{u=0} \atop {s=0}}=0, \ \ \ p,k=0,1,2,... \ \ .
\label{boot}
\end{equation}
The explicit form of the generating function
$\Phi^-_A(u,s)$ is given in the Appendix.

The point of major importance is that, if the calculations are
carried out in the scheme of
{\em renormalized perturbation theory}
with on-shell normalization conditions, the bootstrap equations are
nothing but a system of restrictions for the admissible values of RP's
(real parts of pole positions and triple couplings). In that way
bootstrap system results in a set of constrains for observable
physical spectrum of the theory. Thus once established on the
tree-level, these relations must also hold at higher loop orders, just
because at each loop order one should impose the same RPs. This
explains our direct use of the experimental values of resonance masses
and coupling constants (e.g. given in
\cite{PDG})
to perform the numerical comparison with data. If our scheme is
somehow suitable for the description of physical world the bootstrap
constrains must hold.

\section{Numerical tests}

Now we pass to our numerical tests. As a first example we show how it
is possible to obtain the estimate for the
$G_{\Sigma(1385)\overline{K}N}$
coupling with the help of sum rules that follow from the bootstrap
system. Our first goal is to find the sum rules that can be saturated
with a small number of well established resonances. The up-to-date
information on the
$KN$
resonance spectrum is incomplete in the region of high mass and spin.
Much is unclear with
$M>1 \, GeV$
meson resonances in the
$t$-channel
of elastic reaction. One also needs to keep in mind the possible
existence of
$s$-channel
$S=+1$
exotic resonances. Let us consider the sum rule that follows from the
bootstrap condition
(\ref{boot})
for the invariant amplitude
$A^-$
and corresponds to
$k=p=1$.
It turns out that in this sum rule the contributions of certain not well
established resonances is wiped out. This sum rule can be considered
as purely baryonic one (only baryons with
$J=\frac{3}{2}, \; \frac{5}{2},...$
can contribute), because in the meson sector only isospin
$1$
resonances of odd spin
$J \ge 3$
(e.g.,
$\rho_3(1690)$)
can in principle contribute to it. An assumption is made that heavy
meson contributions are suppressed by small
$\sim \frac{1}{M}$
factors. In our present analysis we also will not take account of
possible contributions of exotic resonances with strangeness
$S=+1$.
However, in what follows we show that several sum rules provide an
indirect evidences in favor of existence of exotic resonances.

Thus we try to saturate our sum rule by the contributions of baryons
with masses
$M < 2 \; GeV$
and spins
$J \le \frac{5}{2}$
(see
\cite{PDG}).
Imputing the deficit to the unknown contribution of
$\Sigma(1385)$
we can estimate the value of
$\Sigma(1385)\overline{K}N$
coupling constant. This gives:
$G_{\Sigma(1385)\overline{K}N}= 1.3 \pm 0.4.$
The experimental value of this constant (see, e.g.,
\cite{Nagels}
p.61)
is:
$
G_{\Sigma(1385)\overline{K}N}=1.06 \pm 0.13.
$
The agreement looks impressive. However, there are several
sufficiently well established resonances with
$M> 2\; GeV$.
The large contribution of
$\Lambda(2100)$
seems to slightly disturb the sum rule. This gives:
$G_{\Sigma(1385)\overline{K}N}= 1.5 \pm 0.7.$
This shift can be compensated by the contributions of
$\Sigma(2100)$
and of the other heavy
$\Sigma$
resonances in this region.

It is very instructive to consider also the sum rules which follow
from the bootstrap constrains for
$A^-$
(\ref{boot})
with many derivatives. These sum rules can be well saturated with the
reliable experimental data on
$S=-1$
baryon spectrum with
$J=\frac{3}{2}, \, \frac{5}{2}$
(mesons with
$J=0,1,2$
do not contribute). This gives a strong evidence in favor of
consistency of our approach, because the shape of these sum rules
crucially depends on our assumptions (in particular, on the concrete
formulation of the
{\em summability principle}
\cite{hsqcd}).
The results are presented in the Table
\ref{table}.
\begin {table}
\begin{center}
\begin{tabular}{||c|c|c||c|c|c||}
\hline
$p $ & $k$ &  Sum  Rule& $p $ & $k$ &  Sum  Rule    \\
\hline
$0$ & $1$ & $13.0 \, \div \, 19.8 \, \simeq \, 19.8 \, \div \, 24.7 $ & $1$ & $1$ & $15.3 \, \div \,  24.2 \, \simeq \, 13.7 \, \div \, 21.8 $ \\
\hline
$0$ & $2$ & $20.7 \, \div \,  25.7 \, \simeq \, 23.4 \, \div \, 28.4 $ &$1$ & $2$ & $16.2 \, \div \, 22.7 \, \simeq \, 14.9 \, \div \, 21.2 $   \\
\hline
$0$ & $3$ & $48.0 \, \div \, 55.1 \, \simeq \, 43.8 \, \div \, 50.9 $& $1$ & $3$ & $23.6 \, \div \, 31.2 \, \simeq \, 23.8 \, \div \, 32.4 $ \\
\hline
$0$ & $4$ & $151.0 \, \div \, 167.3 \, \simeq \, 111.4 \, \div \, 125.1 $ & $1$ & $4$ & $44.0 \, \div \, 55.7 \, \simeq \, 50.2 \, \div \, 66.5 $ \\
\hline
$1$ & $0$  &$23.8 \, \div \, 48.5 \, \simeq \, 24.3 \, \div \, 43.2 $& $1$ & $5$ & $99.8 \, \div \, 123 \, \simeq \, 131.4 \, \div \, 171.8 $\\
\hline
\end{tabular} \; .
\end{center}
\caption{Saturation of sum rules (\ref{boot}) for different values of
$p, \, k$. }
\label{table}
\end{table}
The fact that the balance becomes worse with the growth of
$k$
shows that the contribution of baryons with spin
$J > \frac{5}{2}$
becomes relatively more important in these sum rules.

However, not all sum rules are well saturated with known data. For
example the sum rules for
$A^+$
look very nasty. At first glance, nothing could compensate the huge
positive contribution of
$(I=1, J=\frac{3}{2})$
resonances nearest to the
$KN$
threshold. There are certain possibilities to overcome this
difficulty. First of all, it is interesting to notice that a similar
situation was encountered in the
``toy bootstrap model''
\cite{AVVVKS}
based on Veneziano string amplitude. In certain sum rules for the
resonance parameters of the string amplitude it is sufficient to take
into account the contribution of a relatively small number of first
poles to saturate it with high precision. At the same time, in some
another sum rules it is necessary to sum over the contributions of
considerable number of poles to compensate the
``accidentally large''
contribution coming from several first poles. It is possible that
heavy resonances with
$ J^P=\frac{3}{2}^+ $
and
$ J^P=\frac{5}{2}^+ $
could in principle gradually compensate the large contribution of
$\Sigma(1385)$.
The same mechanism could work for other sum rules from this group with
$k>1$.
Another interesting possibility is to interpret the deficit in these
sum rules as an indirect evidence for the existence of exotic baryons
with strangeness
$S=+1$
(so-called
$Z$ or $\theta$
baryons).
One can easily check that the contribution of a baryon with
$S=+1$
and
$J^P=\frac{3}{2}^+$
below the
$KN$
threshold, or of a
$J^P=\frac{3}{2}^-$
baryon above it, can significantly compensate the deficit. However,
one is forced to assume the existence of at least two exotic baryons
with isospin
$0$
and
$1$,
respectively. Otherwise, it is impossible to attain the mutual
cancellation of the contributions from exotic sector in those sum
rules which are satisfactorily saturated with the
$S=-1$
baryons.

\section{Conclusions}

The numerical tests (that were carried out for
$\pi N$, $KN$, $\pi K$ and $\pi \pi$
reactions) make it possible to conclude that our approach, at least,
does not roughly contradict to presently known phenomenology. However,
at the moment we are unable to give an answer to the main question:
``How many independent RP's are needed to fix the physical content of
effective scattering theory?''
To answer it, we need to somehow solve the bootstrap system. A
possible way to solution is provided by the application of general
theory of analytic continuation along with the tool of
infinite-dimensional matrices. We  also need to study if the higher
order bootstrap constrains
($1$-loop, ...)
impose additional restrictions on the set of RP's or just follow from
the tree-level bootstrap.

\section*{Acknowledgements}
I am grateful to A.~Vereshagin and V.~Vereshagin for multiple help and
advise and to S.~Paston, A.~Vasiliev and M.~Vyazovski for stimulating
discussions.

\section*{Appendix}
Here we give the explicit expressions for the baryon part of the
generating functions of bootstrap system for the amplitude
$A^-$:
$\Phi^-_A(u,s)=
 \sum_{ S=+1 }
\frac{c_I^- G_{R_sKN}F^l_A(-\mathcal{N}M,-(\Sigma+u))}{s-M^2} \, -
\sum_{S=-1}
\frac{b_I^-G_{R_u\overline{K}N}F^l_A(-\mathcal{N}M,-(\Sigma+s))}{u-M^2}.$
The residue of the amplitude in the pole corresponding to a baryon
resonance of strangeness
$S=\pm 1$,
isospin
$I$,
spin
$j=l+\frac{1}{2}$,
normality
$\mathcal{N}$
and mass
$M$
is given by
$
F_A^l(M, \chi)=(M+m)P'_{l+1}(1+\frac{\chi}{2 \phi})+(M-m)
\frac{(M+m)^2-\mu^2}{(M-m)^2-\mu^2}P'_l(1+\frac{\chi}{2 \phi}).
$
Here
$P'_l$
stands for derivatives of ordinary Legender polynomials;
$m \, (\mu)$
is the nucleon (kaon) mass;
$
\phi=\overrightarrow{k}^2_{C.M.F};
$
$
b^-_I, \, c^-_I$
are the isotopic coefficients;
$\Sigma=M^2-2(m^2+\mu^2)$;
$G_{RKN}$
is the dimensionless coupling constant.

\end{document}